\documentclass[prl,reprint]{revtex4-1}

\usepackage{amsmath,amssymb}
\usepackage{verbatim}
\usepackage{graphicx}
\usepackage{hyperref}
\usepackage{color}
\usepackage{mathrsfs}

\usepackage{amsthm}

\usepackage{amsfonts}

\theoremstyle{definition}

\newcommand{\NN}{\mathbb{N}} 
\renewcommand{\emptyset}{\varnothing} 

\newcommand{\Hi}{\mathcal{H}} 

\newcommand{\M}{{\mathcal M}} 
\newcommand{\be}{\begin{equation}}
\newcommand{\ee}{\end{equation}}
\newcommand{\bea}{\begin{eqnarray}}
	\newcommand{\eea}{\end{eqnarray}}
\newcommand{\non}{\nonumber}
\newcommand{\ba}{\begin{array}}
	\newcommand{\ea}{\end{array}}

\def\1{\frak 1}
\def\2{\frak 2}
\def\3{\frak 3}

\newcommand{\OO}{\mathcal{O}}

\newlength{\oldcolsep}

\begin{document}

\title{Quantum Mechanics from Ergodic Average of Microstates}

\author{Marco Matone}

\affiliation{Dipartimento di Fisica e Astronomia ``G. Galilei'' \\
 Istituto
Nazionale di Fisica Nucleare \\
Universit\`a di Padova, Via Marzolo, 8-35131 Padova,
Italy}


\date{\today}

\begin{abstract}

We formulate quantum mechanics as an effective theory of an underlying structure characterized by microstates $|\M^j(t)\rangle$, each one
defined by the quantum state $|\Psi(t)\rangle$ and a complete set of commutative observables $O^j$. At any time $t$,
$|\M^j(t)\rangle$ corresponds to a state $|O^j_k\rangle$, for some $k$ depending on $t$, and jumps after time intervals 
whose duration, of the order of
the Compton time $\tau$, is proportional to the probability $|\langle O_k^j|\Psi(t)\rangle|^2$. This reproduces the Born rule and mimics the wave-particle duality.
The theory is based on a partition of time whose flow is characterized by quantum probabilities.
Ergodicity arises at ordinary quantum scales with the expectation values corresponding to time averaging over a period $\tau$. The measurement of $O^j$ provides a new partition of time
and the outcome is the state $|O_k^j\rangle$ to which $|\M^j(t)\rangle$ corresponds at that time.
The formulation, that shares some features with the path integral, can be tested by experiments involving time intervals of order $\tau$.

\end{abstract}

\maketitle
\allowdisplaybreaks

\textit{Introduction and summary.}---We show that quantum mechanics can be formulated in terms of ergodic average of microstates.
The key point is that physical states have determined quantum numbers for time intervals which are of the order of Compton time
\be
\tau={ h\over mc^2} \ .
\ee
The role of the Schr\"odinger equation is to determine a partition of time by fixing the evolution of probabilities.
A quantum state $|\Psi(t)\rangle$ and a complete set of commuting observables (CSCO) $O^j$ define a microstate $|\M^j(t)\rangle$
describing $O^j$ with defined values and jumping at given times.
The set of probabilities associated to the eigenvalues $O_k^j$'s is proportional to the time interval during which $|\M^j(t)\rangle$
corresponds to $|O_k^j\rangle$.
At the quantum scales such jumps mimic the wave-particle duality behavior.

The formulation reproduces the results of quantum mechanics, in particular the Born rule, with corrections of the order of $\tau$
that may be experimentally tested.
The expectation value of the observables corresponds to the time averaging over a cycle of length $\tau$. The outcome of a measurement of
$O^j$ at time $t$ is just the eigenvalue of the self-adjoint operator $\hat O^j$ acting on $|\M^j(t)\rangle$.

Some features of the formulation resemble the intermediate time partition \cite{Dirac:1933xn} in the Dirac-Feynman path
integral \cite{Dirac:1933xn}\cite{Feynman:1948ur}.

\textit{Time partition by quantum probabilities.}---Let $|\Psi(t)\rangle$ be a normalized quantum state in a Hilbert space $\Hi$. Set
\be
\Delta_N:=(N\tau,(N+1)\tau] \ ,
\ee
with $N=[t/\tau]$ the integer part of $t/\tau$.
In the following, for any $N$, we make a partition of $\Delta_N$ by the probabilities $|\langle O_k|\Psi(t)\rangle|^2$,
with $\{O_k\}$ the spectrum of the self-adjoint operator $\hat O$ associated to the observable $O$.

For any $O_k$ and $N$, fix a set of $2n_k$ ordered times
\be
N\tau\leq t_{k,1}(N)<\ldots <t_{k,2n_k}(N)\leq(N+1)\tau \ ,
\label{ordering}\ee
satisfying the condition
\be
\sum_{j=1}^{n_k}(t_{k,2j}(N)-t_{k,2j-1}(N))=|\langle O_k|\Psi(N\tau)\rangle|^2\tau \ .
\label{thekey}\ee
Define, for $j\in[1,n_k]$, the sets with empty intersection
\be
I_{k,j}(N)=(t_{k,2j-1}(N),t_{k,2j}(N)] \ .
\ee
We require empty intersection even for the time intervals associated to each pair $O_j,O_k$, $j\neq k$,
that is
\be
I_{j,l}(N)\cap I_{k,m}(N)=\emptyset \ .
\label{setA}\ee
Consider the subset of $\Delta_N$
\be
\{|\Psi(t) \rangle,O_k\} := I_{k,1}(N)\cup  \ldots \cup I_{k,2n_k}(N)  \ ,
\label{theinitial}\ee
and note that (\ref{ordering}), (\ref{thekey}),  (\ref{setA}) and the condition
\be
\sum_k|\langle O_k|\Psi(t)\rangle|^2=1 \ ,
\label{unitarity}\ee
imply that the $t_{k,j}(N)$'s also satisfy the relation
\be
\bigcup_k \{|\Psi(t)\rangle,O_k\}=\Delta_N \ .
\label{set}\ee
The above construction provides a partition of time that depends on
$|\Psi(t)\rangle$ and $\{O_k\}$.
This means that the quantum probabilities, seen as independent data, constrain the events and therefore the time flow.
In the case of a time independent Hamiltonian $H$, the coefficient $|\langle O|\Psi(t)\rangle|$ is time independent when $[H,\hat O]=0$.
It follows that the times $t_{k,j}(N)$'s associated to the observables commuting with a time independent Hamiltonian can be chosen in such a way that
\be
t_{k,j}(N)=N\tau+t_{k,j}(0) \ .
\ee
In this case the length of the intervals $I_{k,j}$ is independent of $N$. It follows that such observables, and in particular the Hamiltonian,
play a special role as each eigenvalue $O_k$ defines $n_k$ time intervals of period $\tau$.

\textit{Time and observables.}---Here the Schr\"odinger equation describes the evolution
of time partition. A related quantity is the step function
\be
S_k(t) := \begin{cases} 1\; \textrm{if}\; t\in \{|\Psi(t)\rangle,O_k\} \ , \\
0\; \textrm{if}\; t\notin \{|\Psi(t)\rangle,O_k\} \ .
\end{cases}
\label{onekey}\ee
that satisfies remarkable properties. First, we have
\be
< S_k(t)>_{\tau}=|\langle O_k|\Psi(N\tau)\rangle|^2 \ ,
\label{expect}\ee
where $<\cdot>_\tau$ denotes the time average
\be
< f(t) >_{\tau}:={1\over\tau}\int_{\Delta_N}dt f(t) \ .
\ee
Next, (\ref{setA}) and (\ref{set}) imply that for all $t>0$
there is one and only one value of $k$ such that $S_k(t)\neq0$. Hence,
\be
\sum_k S_k(t)=1 \ .
\label{uno}\ee
For all $t>0$ and $j,k$, we have the key relation
\be
S_j(t)S_k(t)=\delta_{jk}S_j(t) \ ,
\label{idempotent}\ee
that can be interpreted as an intriguing orthonormality condition with indexes $j,k$ and $t$.

We associate to each $O_k$ a function of $t$ which takes the value $O_k$ in $I_{k,j}(N)$ if $t\in I_{k,j}(N)$ and 0 otherwise
\be
{\OO}_k(t):=S_k(t)O_k \ ,
\label{determine}\ee
and set
\be
{\OO}(t)=\sum_k {\OO}_k(t) \ .
\label{sixteen}\ee
We have $\hat\OO(t)|O_k\rangle = \OO_k(t) |O_k\rangle$, where
\be
\hat\OO(t):=\sum_k {\OO}_k(t)|O_k\rangle\langle O_k| \ .
\ee
Note that for $t\notin \{|\Psi(t) \rangle,O_k\}$, $\hat\OO(t)$ acts as annihilator operator
$\hat\OO(t)|O_k\rangle=0$.
By (\ref{expect}) the state
\be
|O_k(t)\rangle:= S_k(t)|O_k\rangle \ ,
\ee
has time average
\be
< |O_k(t)\rangle >_{\tau} = |\langle O_k|\Psi(N\tau)\rangle|^2|O_k\rangle \ .
\ee
$|O_k(t)\rangle$ is eigenstate of $\hat O$ with eigenvalues that may be either $O_k$ or $0$. Furthermore,
by (\ref{idempotent}), $|O_k(t)\rangle$ is eigenstate
even of $\hat\OO(t)$
\be
\hat\OO(t)|O_k(t)\rangle=\OO_k(t)|O_k(t)\rangle \ .
\ee
Note that $\langle O_j(t)|O_k(t)\rangle=\delta_{jk}S_j(t)$.
For any $t$, $\{|O_k(t)\rangle\}$ describes a one dimensional subspace $\Hi(t)$ of $\Hi$ that in
the time $\Delta_N$ spans the subspace $\Hi_N$ of $\Hi$, with basis
\be
\bigotimes_{k} <|O_k(t)\rangle>_{\tau} = \bigotimes_{k}|\langle O_k|\Psi(N\tau)\rangle|^2|O_k\rangle \ .
\ee

Let us give an example of how $|\Psi(t)\rangle$ and $O$ define each $S_k(t)$, and therefore ${\OO}(t)$, by considering the case in which
the unique non vanishing probabilities are
\be
l(t):=|\langle O_1|\Psi(t)\rangle|^2 \ , \quad |\langle O_2|\Psi(t)\rangle|^2=1-l(t) \ .
\ee
Choose $n_1=1$, $n_2=2$. Eq.(\ref{ordering}) implies $t_{k,1}(N)=N\tau$. $k=1$ would imply
$t_{2,2}(N)=t_{2,3}(N)$, so that $k=2$. By (\ref{set}) and (\ref{thekey})
$t_{2,2}(N)=t_{1,1}(N)$ and $t_{1,2}(N)=t_{2,3}(N)=t_{1,1}(N)+\tau l(N\tau)$. Finally $t_{2,4}=(N+1)\tau$.
Therefore,
\be
\OO(t)=\begin{cases} O_1\; \textrm{if}\; t\in I_{1,1}  \ , \\
O_2\; \textrm{if}\; t\in I_{2,1}\cup I_{2,2} \ ,
\end{cases}
\ee
where
\bea
I_{2,1} & = & (N\tau ,t_{11}(N)] \ , \non \\
I_{1,1} & = & (t_{1,1}(N),t_{1,1}(N)+\tau l(N\tau)] \ , \\
I_{2,2} & = & (t_{1,1}(N)+\tau l(N\tau),(N+1)\tau] \ . \non
\eea

\textit{Microstates and Born rule.}---In the following, we will associate the probability of finding the value $O_k^j$ in a measurement of $O^j$ at a random time $t\in\Delta_N$
 to the sum of the time intervals in $\Delta_N$ associated to $O_k^j$. To this end, we first generalize the previous
 construction to the set of all observables $\{O^j\}$.
Denote by $t_{k,l}^j(N)$ the times associated
to each eigenvalue $O_k^j$ of $\hat O^j$, defined in the same way as in the case of $O_k$. The associate time intervals have one more
index labeling the different observables
\be
I_{k,l}^j(N)=(t_{k,2l-1}(N),t_{k,2l}^j(N)] \ .
\ee
Furthermore, the step function now is $S_k^j(t)$, defined as in (\ref{onekey}) but with $O_k$ replaced by $O_k^j$.

The new ingredient is the relation between the time intervals corresponding to each pair of observables, $O^j$ and $O^k$.
Let us first investigate the case of times associated to a CSCO. We consider the example of the basis
state $|E_n,l,m\rangle$ in the Hilbert space of the hydrogen atom.
As in (\ref{unitarity}), for any possible $|\Psi(t)\rangle$, we have $\sum_{n,l,m}|\langle E_n,l,m|\Psi(t)\rangle|^2=1$.
Therefore, as far as the time partition is concerned, $E$, $L^2$ and $L_z$ can be considered as a single observable $O$,
with the corresponding eigenvalues labeled by three indexes, that is $O_{n,l,m}$. As in the case of
a single observable, for each $|\Psi(t)\rangle$, even $O_{n,l,m}$ provides a
partition of time. This implies that, for any $t$, $\OO(t)$ corresponds to
$O_{n,l,m}$, with $n,l,m$ fixed by $t\in\{|\Psi(t)\rangle,O_{n,l,m}\}$. That is, the three observables have, simultaneously and for the
time intervals $\{|\Psi(t)\rangle,O_{n,l,m}\}$, the values $E_n$, $l$ and $m$.
Such a construction extends to all CSCO's.
In the following we denote a CSCO by $O^j$ with fixed $j$. The index of an eigenvalue will be a multi-index, e.g. $O_k^1=\{E,L^2,L_z;{n,l,m}\}$.

For each CSCO $O^j$ we define the {\it microstate}
\be
|\M^j(t)\rangle:=\sum_l |O_l^j(t)\rangle \ ,
\label{mic}\ee
By  $\langle O_k^j|\M^j(t)\rangle=S_k^j(t)$ it follows that the state
$|\M^j(t)\rangle$ jumps instantaneously from an element of the basis $\{|O_k^j\rangle\}$ of $\Hi$ to another one. In particular,
\be
|\M^j(t)\rangle = |O_k^j\rangle \ ,
\ee
with $k$ such that $t\in\{|\Psi(t)\rangle,O_k^j\}$. Also note that $|\M^j(t)\rangle$ is eigenstate of $\hat\OO^j(t)$
with eigenvalue $\OO^j(t)=\sum_kS_k^j(t)O_k^j$
\be
\hat\OO^j(t)|\M^j(t)\rangle=\OO^j(t)|\M^j(t)\rangle \ .
\label{eigenstate}\ee
Furthermore, for any $t>0$, (\ref{uno}) implies
\be
\sum_k\langle O_k^j|\M^j(t)\rangle=1 \ ,
\ee
and by (\ref{idempotent})
\be
\langle O_k^j|\M^j(t)\rangle \langle O_l^j|\M^j(t)\rangle= \delta_{kl}\langle O_k^j|\M^j(t)\rangle \ .
\ee
Finally, for any $t>0$
$\langle\M^j(t)|\M^j(t)\rangle=1$.

By construction, each eigenstate $|O_k^j\rangle$ in the microstate (\ref{mic}) appears for a time interval
which is proportional to the probability of finding the eigenvalue $O_k^j$ in a measurement of $O^j$ in the state $|\Psi(t)\rangle$.
As we will see in more detail below, if the
state is described by $|\M^j(t)\rangle$ then a measurement at random times
of $O^j$ reproduces the probability distribution provided by the rules of standard quantum mechanics.

Let us now consider the case of two non commutative observables, $O^j$ and $O^k$. Such observables define partition
of times which are incompatible with the mentioned correspondence between
the probability distributions in a measurement of $O^j$ and $O^k$. The point is that a partition of time intervals requires that at each time
a given observable has fixed values. Therefore, for any $t$, the eigenvalue $O_k^j$ may be or not the corresponding value of $O^j$. The answer
is yes or not, and corresponds to the orthonormality condition $\langle O_l^j|O_k^j\rangle=\delta_{lk}$.
On the other hand, expanding the elements of the basis $|O_l^j\rangle$ in (\ref{mic}) in terms of the basis $|O_m^k\rangle$ one sees that, because of
the difference between $S_l^j(t)$ and $S_m^k(t)$, the resulting expression of $|\M^j(t)\rangle$
cannot correspond to the one of $|\M^k(t)\rangle$. In other words, for any $|\Psi(t\rangle$,
non commutative sets of CSCO, and therefore the associated bases in the Hilbert space, define different partitions of time. This is how direct simultaneous
measurement  of noncommutative pairs of CSCO, and therefore even of single non commutative observables, are excluded in the present formulation.

A property of microstates is that the probability of finding
the value $O_k^j$ in a measurement of $O^j$ in  $|\M^j(t)\rangle$ at a random time $t\in\Delta_N$ is proportional to the sum of the time intervals
between $N\tau$ and $(N+1)\tau$. In other words, by (\ref{thekey}), we recover the Born rule
\bea
P_{N\tau}(O_k^j) & = &
{1\over\tau}\sum_{l=1}^{n_k}(t_{k,2l}^j(N)-t_{k,2l-1}^j(N)) \non \\
& = & |\langle O_k^j|\Psi(N\tau)\rangle|^2  \ ,
\label{almost}\eea
that can be also expressed in the form
\be
P_{N\tau}(O_k^j)=<\langle O_k^j|\M^j(t)\rangle>_{\tau}=<|\langle O_k^j|\M^j(t)\rangle|^2>_{\tau} \ .
\label{withoutinterference}\ee
The last equality, consequence of (\ref{idempotent}),
is reminiscent of the standard quantum formula. The reason why the interference terms do not appear
in the square in (\ref{withoutinterference}) is that the time partition induces the orthonormal property (\ref{idempotent}). On the other hand,
even if $|\M^j(t)\rangle$ depends only on the modulus of the $\langle O_k^j|\Psi(t)\rangle$'s, the full information in
the state $|\Psi(t)\rangle$
can be reobtained from the structure of the set $\{|\M^j(t)\rangle\}$. In particular,
the full set $\{\langle O_k^j|\Psi(t)\rangle; k\in I\}$ can be reconstructed from the set $\{|\langle O_k^j|\Psi(t)\rangle|; k\in I, j\in J\}$.

The difference between $P_{N\tau}(O_k^j)$ and the standard value, that is $|\langle O_k^j|\Psi(t)\rangle|^2$,
is of order $t-N\tau$. In this respect, it is worth mentioning that at the present the most accurate measure of time in a quantum experiment concerns the
absolute zero of time in the photoelectric effect.
This has been measured to a precision better than 1/25th of the atomic unit of time, that is $\approx 10^{-18}$s \cite{attosecond}. This suggests
that in a near future it will be possible to consider experiments measuring scales of the Compton time of the electron $\approx 10^{-20}$s. A possible experimental
test of the formulation is to perform multiple measurements of the electron spin of identical prepared states, at time intervals shorter than the Compton time, and checking
a deviation from the probability distribution due to the granularity of the time partition.
We also note that neutrinos have a Compton time which is much bigger than the one of the electron, therefore the jumps of their quantum numbers should be more easily observed.
This may shed light on their oscillations.

It is immediate to check that
\be
<\OO^j(t)>_{\tau} = \sum_k|\langle O_k^j|\Psi(N\tau)\rangle|^2 O_k^j \ ,
\label{timeaverage}\ee
so that we have that at ordinary quantum scales
\be
<\OO^j(t)>_{\tau}=\langle \Psi(t)|\hat O^j|\Psi(t)\rangle \ .
\ee
Note that integrating $ \OO^j(t)$ between $\alpha\tau$ and $(\alpha+1)\tau$, with $\alpha\notin\NN$, would give, in general,
a different value with respect to the case $\alpha\in \NN$.
The reason is that, as we said,
periodicity of the time intervals can be chosen if the $|\langle O_k^j|\Psi(t)\rangle|$'s are time independent. This means that,
besides (\ref{timeaverage}), the quantum mean
values can be obtained by averaging over larger time intervals. The difference remains extremely small.

\textit{Evolution of microstates after a measurement.}---In the present formulation the outcome of a measurement of an observable at time $t$ is the
value of such an observable in the corresponding microstate at that time. As we said, such a value is always defined, namely
$|\M^j(t)\rangle=|O_k^j\rangle$ with $k$ such that $t\in\{|\Psi(t)\rangle,O_k^j\}$. The key point is that the physical state is
represented by the set $\{|\M^j(t)\rangle\}$ rather than
$|\Psi(t)\rangle$. This means that, due to the different time partitions, each CSCO has a different time evolution, so that
the outcome of a measurement of the CSCO $O^j$ is equivalent to act with $|O^j\rangle\langle O^j|$ on the corresponding state $|\M^j(t)\rangle$.
The state of a physical system is then represented by
\be
|\M(t)\rangle:=|\M^1(t), \ldots, \M^n(t)\rangle \ .
\ee
The effect of the measurement of $O^j$ at the time $t_0$ is to
replace, for $t\geq t_0$, $|\Psi(t)\rangle$ by
\be
|\Psi(t)\rangle= U(t,t_0)|\M^j(t_0)\rangle \ ,
\label{newPsi}\ee with $U(t,t_0)$
the time evolution operator.
This has some analogy with the measurement of a given physical quantity of a classical object, such an operation is a perturbation and then the future dynamics
is determined by reconsidering the equations of motion but with the new initial data given by the outcome of the measurement. If the measurement of $O^j$ has given the value $O_k^j$,
then there is a resetting of the time partition, fixed by the new $|\Psi(t)\rangle$ in (\ref{newPsi}), associated to each one of the CSCO $O^1,\ldots,O^n$.
By  (\ref{determine}) this fixes, for $t\geq t_0$, the set $\{S_k^j(t)\}$ and therefore $|\M(t)\rangle$. Note that each CSCO has determined
values but, due to the illustrated mechanism, the outcome concerning a measurement of $O^j$ followed by $O^k$ is in general different if the measurement is
done in the reverse order. It should be also observed that
even if the $2n_l^j$ and $2n_m^k$ times $t_{l,p}^j$'s and $t_{m,q}^k$'s are constrained by the probabilities conditions, we do not know all their values.

We note that the outcome of a measurement can be also described by acting on the quantum state $|\Psi(t)\rangle$. This follows immediately by introducing the the measurement operator
\be
\hat M^j(t)=\sum_k |O_k^j(t)\rangle\langle O_k^j(t)|=\sum_k S_k^j(t)|O_k^j\rangle\langle O_k^j| \ ,
\ee
whose action on $|\Psi(t)\rangle$ is
\be
\hat M^j(t)|\Psi(t)\rangle= \langle O_k^j|\Psi(t)\rangle|O_k^j\rangle \ ,
\ee
with $k$ such that $t\in\{|\Psi(t)\rangle,O_k^j\}$.

\textit{Analogies with the path integral.}---In \cite{Dirac:1933xn} Dirac noticed that an operatorial analogue
of the Hamilton-Jacobi relations $p=\partial_qS$, $P=-\partial_Q S$, with $S(q,Q,t)$ the Hamilton principal
function, follows by the correspondence
\be
\langle q+\delta q,t+\delta t|q, t\rangle \sim \exp({i\over\hbar}\delta t L) \ .
\label{DiracFundamentalIdea}\ee
He then introduced the partition of the time interval $t-t_0$ into $n$ infinitesimal
intervals $t_k=t_0+n\epsilon$, $t_n=t$ and used the completeness relation to get \cite{Dirac:1933xn}
\be
\langle q,t|q_0,t_0\rangle=\int \prod_k dq_k \langle q,t|q_{n-1},t_{n-1}\rangle \cdots \langle q_1,t_1| q_0,t_0\rangle  \ .
\ee
Taking the limit $\epsilon\to0$ with $n\epsilon$ fixed, leads to the Dirac-Feynman path integral
\cite{Dirac:1933xn}\cite{Feynman:1948ur}
\be
\langle q,t|q_0,t_0\rangle =\int Dq e^{{i\over\hbar}\int_{t_0}^{t} dt' L(q,\dot q)} \ .
\label{pathintegral}\ee
Noticing that at time $t_0$ the state corresponds to
$\psi_{q_0}(q,t_0)=\delta(q-q_0)$, one sees that the wave functional
\be
\psi_{q_0}(q,t)=U(t,t_0) \delta(q-q_0) \ ,
\ee
is identical to $\langle q,t|q_0,t_0\rangle$. Equivalently, we have  $|\Psi(t)\rangle=U(t,t_0)|q_0,t_0\rangle$, so that
$\langle q,t|q_0,t_0\rangle=\langle q|\Psi(t)\rangle$. 
In our formulation, we are interested in $|\langle q|\Psi(t)\rangle|^2$.

Now note that the above time partition considered by Dirac, also induces a partition of the $q$-space. This is in the same spirit of our construction and suggests considering a modification
of the path integral by considering the position $q$ as having a discrete spectrum
\be
\hat q |q_k\rangle=q_k|q_k\rangle \ .
\ee

Space-time discretization should arise at the scales where gravity starts playing the key role and where the notions
of causality and distance between events loose their standard meaning, that is at the scales of the Planck length  $\ell_P=\sqrt{\hbar G/c^3}$. We then consider the positions of the particle the set $\{q_k\}$, with
$q_{k+1}-q_k=\ell_P$ and define
\be
{\rm Pr}(q_k):= \int_{q_k}^{q_{k+1}}dq |\langle q| \tilde\Psi(t)\rangle|^2 \ ,
\ee
where
\be
| \tilde\Psi(t)\rangle= \Bigg[\int_{\bar q_k-\lambda/2}^{\bar q_k+\lambda/2}dq |\langle q |\Psi(t)\rangle|^2\Bigg]^{-1/2}|\Psi(t)\rangle \ ,
\ee
with  $\bar q_k=q_k+\ell/2$ and $\lambda=h/mc$ the Compton wavelength. One then can repeat the above construction even for the position $q$ by considering the time partition
associated to each $q_k$, with $|\langle O_k^j|\Psi(t)\rangle|^2$ replaced by ${\rm Pr}(q_k)$.

\textit{Acknowledgements.}---We thank P.~A.~Marchetti and R.~Volpato for useful comments and
discussions.


\begin{thebibliography}{99}


\bibitem{Dirac:1933xn}
  P.~A.~M.~Dirac,
  Phys.\ Z.\ Sowjetunion {\bf 3}, 64  (1933).

\bibitem{Feynman:1948ur}
  R.~P.~Feynman,
  Rev.\ Mod.\ Phys.\  {\bf 20},  367 (1948).

\bibitem{attosecond}
  M. Ossiander {\it et al.},
  Nature Phys. {\bf 13}, 280 (2017). 


\end{thebibliography}
\end{document}